\documentclass[conference]{IEEEtran}
\IEEEoverridecommandlockouts
\usepackage[hidelinks]{hyperref}
\usepackage[numbers,sort&compress]{natbib}
\usepackage{amsmath,amssymb,amsfonts}
\usepackage{algorithmic}
\usepackage{graphicx}
\usepackage{textcomp}
\usepackage{xcolor}
\usepackage{listings}
\usepackage{caption}
\usepackage{subcaption}
\usepackage{adjustbox}
\usepackage{booktabs}
\usepackage{float}
\usepackage{multirow}
\usepackage{threeparttable}

\usepackage{fancyhdr}

\makeatletter
\let\ps@IEEEtitlepagestyleORIG\ps@IEEEtitlepagestyle
\def\ps@IEEEtitlepagestyle{%
  \ps@IEEEtitlepagestyleORIG
}
\makeatother

\def\BibTeX{{\rm B\kern-.05em{\sc i\kern-.025em b}\kern-.08em
    T\kern-.1667em\lower.7ex\hbox{E}\kern-.125emX}}
\begin{document}

\title{A Framework for Accelerating Transformer Inference on RISC-V for Edge AI\\
\thanks{This work was supported by Taighde Éireann – Research Ireland (18/CRT/6183, 22/US/3848), the National Science Foundation (2315851, 2106634), the Department for the Economy, Northern Ireland (USI-226), and the Microelectronic Circuit Centre Ireland (MCCI-2020-07).}

}

\author{
\IEEEauthorblockN{
Ajay Kumar M\IEEEauthorrefmark{1},
Vishnu PS\IEEEauthorrefmark{1},
Yike Li\IEEEauthorrefmark{1},
Shreejith Shanker\IEEEauthorrefmark{2}\\
Dimitrios S. Nikolopoulos\IEEEauthorrefmark{3},
Bo Ji\IEEEauthorrefmark{3},
Hans Vandierendonck\IEEEauthorrefmark{4},
Deepu John\IEEEauthorrefmark{1}
}

\IEEEauthorblockA{\IEEEauthorrefmark{1}University College Dublin, Ireland\hspace{1em}%
\IEEEauthorrefmark{2}Trinity College Dublin, Ireland\\%
\IEEEauthorrefmark{3}Virginia Tech, USA\hspace{1em}%
\IEEEauthorrefmark{4}Queen's University Belfast, UK}
}


\maketitle
\begin{abstract}
This work presents a framework for accelerating transformer-based language models (LMs) on resource-constrained IoT devices. The framework targets compact LMs: BERT-Tiny (B-Ty), MobileBERT (M-Bt), MiniLM (M-Lm), Electra (E-Lt) and DeBERTa (D-Bt) — selected for their architectural diversity and use in edge inference scenarios. The proposed flow derives lightweight instruction set extensions tailored to the non-obvious computational patterns of these models. In addition, a custom instruction is introduced to accelerate the address generation stage of batch matrix multiplication, achieving a performance improvement of 15.39–21.74\% with modest ASIC overheads of 6.79\% in area and 2.33\% in power. To further enhance performance without incurring additional processor core hardware cost, an optional compiler-directed loop unrolling strategy is employed, trading increased code size for overall reduced execution time. Evaluation on the Synopsys trv32p3f RISC-V core demonstrates inference speedups of up to 2.19×, and FPGA implementation on the AMD Zynq UltraScale+ ZCU102 shows a 32.94\% area overhead at 75 MHz, whereas the ASIC implementation using the TSMC  28 nm library incurs a 21.08\% area overhead while operating at 250 MHz.

\end{abstract}


\section{Introduction}


Deploying deep neural networks (DNNs) on bare-metal IoT platforms remains difficult under tight energy, memory, and compute budgets. Many acceleration frameworks still depend on heavyweight stacks, or lack fine-grained architectural customisation, hindering practical deployment in deeply embedded settings. As modern edge workloads increasingly incorporate transformer-based architectures, particularly in vision\cite{saha2025vision} and language tasks\cite{zheng2025review}, there is a growing need for model class-specific optimisations targeting these networks. Recent efficient edge Transformers and LMs\cite{bhargava2021generalization, DBLP:journals/corr/abs-1908-08962, sun2020mobilebert, 10.5555/3495724.3496209, clark2020electra, he2023debertav} still present distinct computational characteristics. The higher reliance on attention patterns and more extensive use of matrix multiplications introduce new challenges in data movement for existing instruction-level optimizations designed for CNN workloads, which leads to inefficient performance on edge platforms.

Prior work in vision~\cite{min2023energy}, NLP~\cite{9658643, 11038291, 10.1145/3719276.3727954}, and general DNN acceleration~\cite{11235344, 10.1145/3665283.3665342} shows that pruning, quantization, ISA extensions, and dedicated acceleration units can enable efficient RISC-V deployment. \cite{min2023energy} accelerates a pruned ViT via specialized sparse-load instructions. RISC-VTF~\cite{9658643} adds custom matrix and activation ops. VEXP~\cite{11038291} couples ISA extensions with arithmetic blocks for fast, accurate softmax. However, existing works do not address the limited compiler support for the proposed extensions, often requiring users to generate manual assembly implementations to extract performance gains. V-SEEK~\cite{10.1145/3719276.3727954} emphasizes quantization and toolchains, but primarily evaluates throughput/latency, without discussion of power consumption, despite its importance as a key metric in resource-constrained edge devices. ~\cite{suarez2025energy} further points out that the effectiveness of front-end optimizations like quantization is highly dependent on the underlying architecture, motivating the need for an end-to-end profiling and optimization framework through HW/SW co-design and co-optimization. 

\par In this work, instead of targeting vector extensions on dedicated compute components, we focus on scalar extensions that can execute solely on a standalone RISC-V core. By avoiding the area and power-intensive overhead of vector units, our approach reduces hardware cost, preserves ease of compiler integration and programmability, and enables practical acceleration of LMs under strict edge constraints. The key contributions of this work are as follows:

\begin{enumerate}
    \item A framework for model-class-aware RISC-V ISA extension generation targeting compact transformer LMs.

    \item A custom fused instruction accelerating batch matrix multiplication with minimal area/power overhead.

    \item An optional compiler-directed loop-unrolling strategy  that improves inference throughput without incurring additional hardware resources.
\end{enumerate}

\section{Methodology}

\begin{figure*}
    \centering
    \includegraphics[height=2cm, width=1\linewidth]{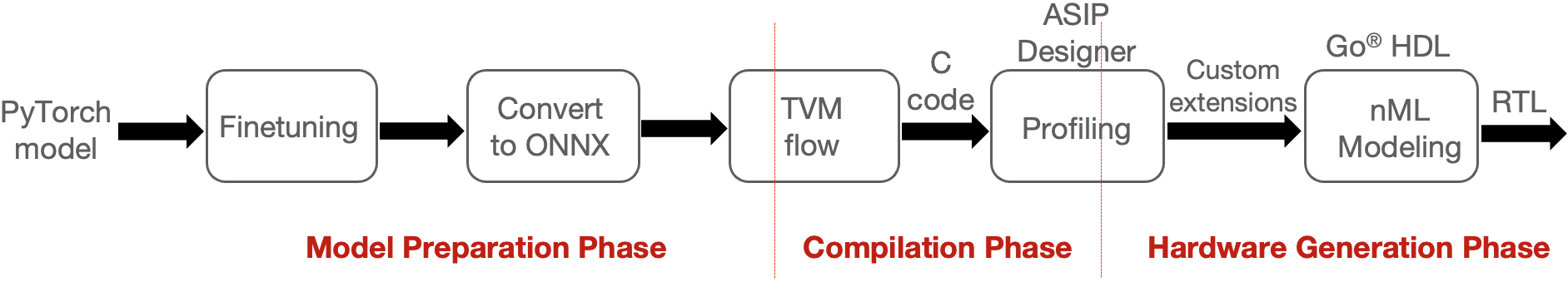}
    \caption{The proposed flow processes a PyTorch model to generate custom instructions for a class of similar models.}
    \label{high_level_flow}
    \vspace{-2mm}
\end{figure*}

The trv32p3f processor core from Synopsys is used as a baseline. It has a 32-bit wide data path and a three-stage pipeline. This core implements the RV32IMFZfinx ISA and has floating-point hardware support. The extended core is tested with variants of standard language models' C codes. Fig.~\ref{high_level_flow} outlines the high-level flow, which is detailed in the following sub-sections.

\begin{figure*}
  \centering
  \begin{minipage}{0.19\textwidth}
    \centering
    \includegraphics[height=2.8cm]{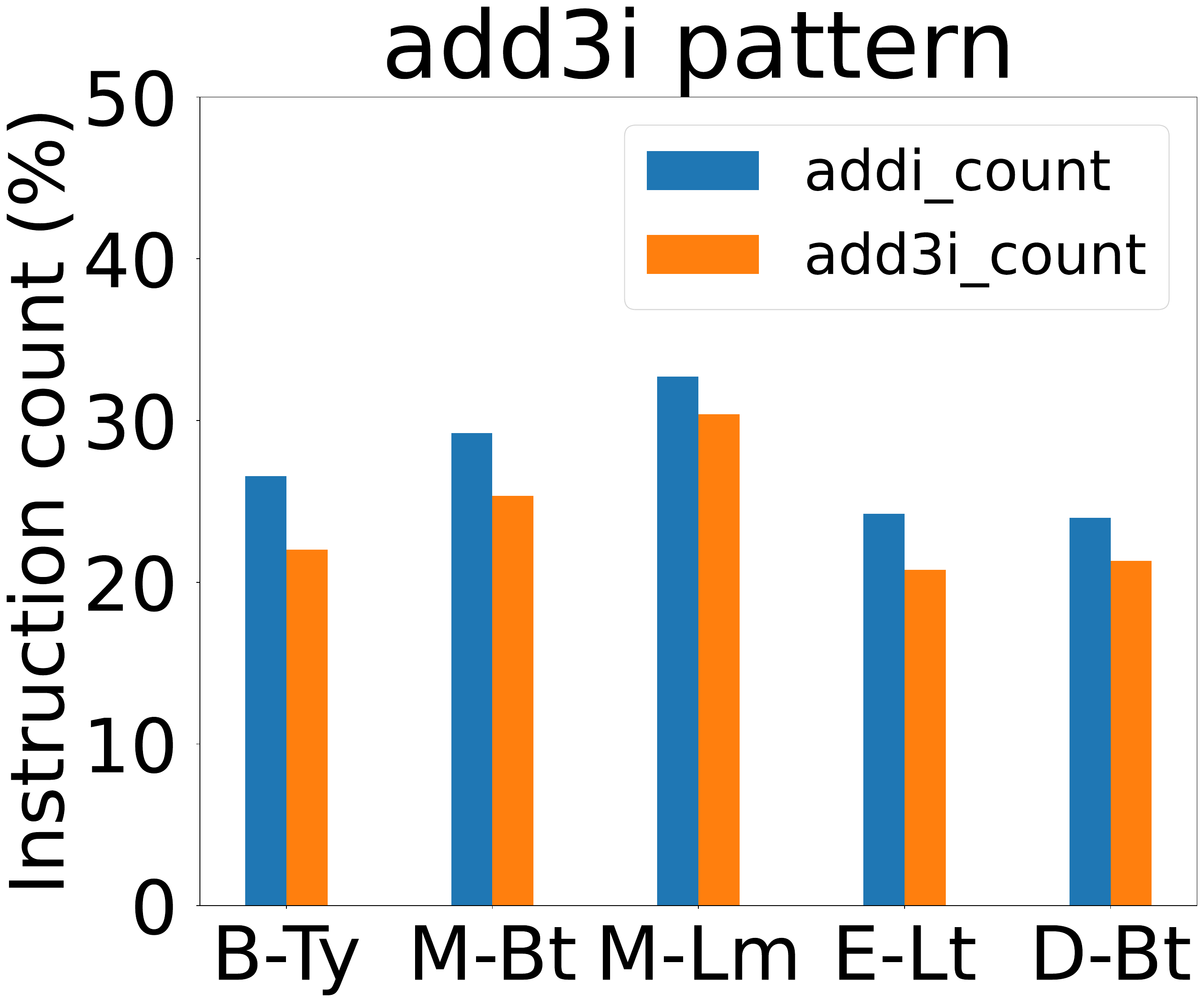}
  \end{minipage}%
  \begin{minipage}{0.19\textwidth}
    \centering
    \includegraphics[height=2.8cm]{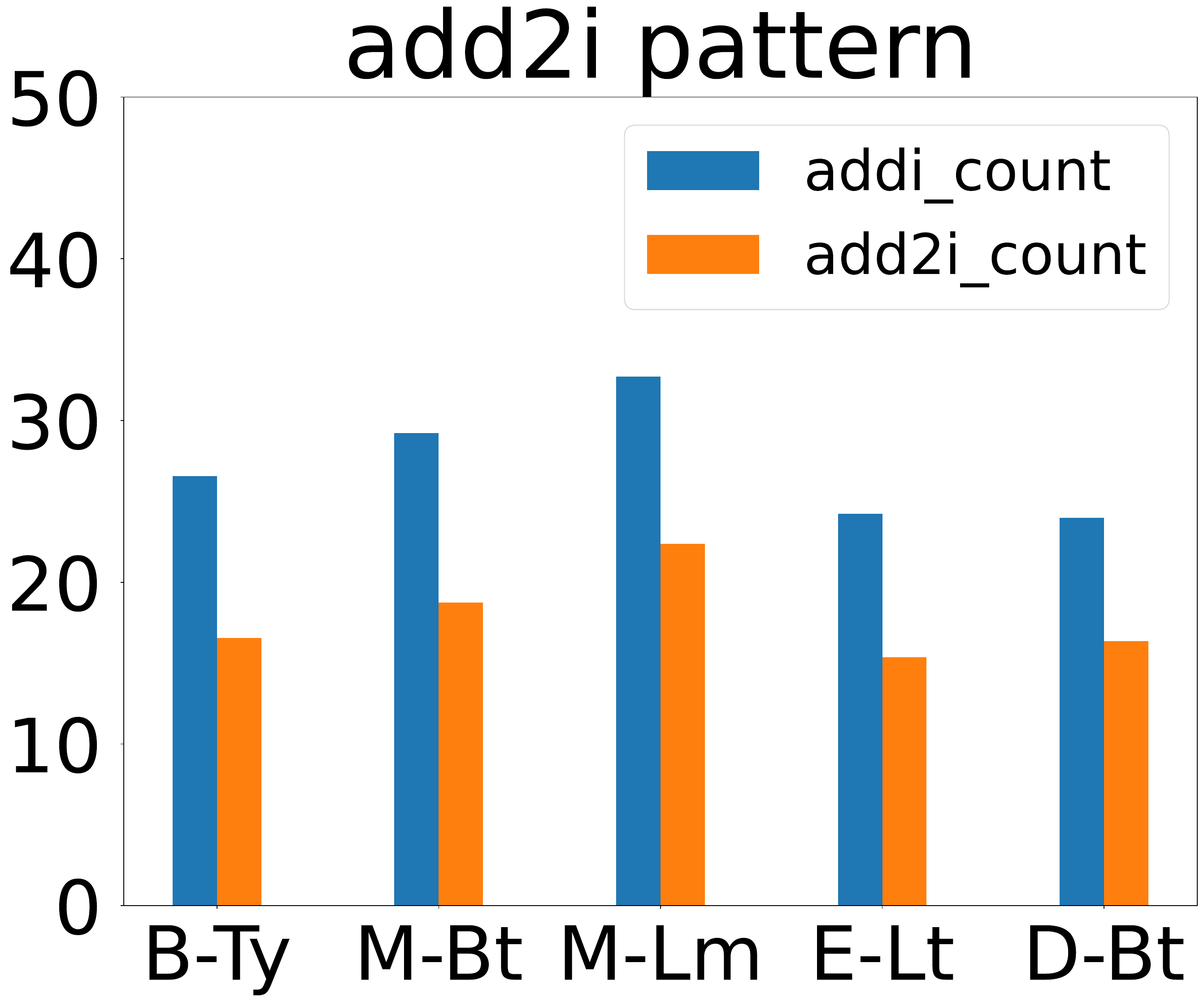}
  \end{minipage}
    \begin{minipage}{0.19\textwidth}
    \centering
    \includegraphics[height=2.8cm]{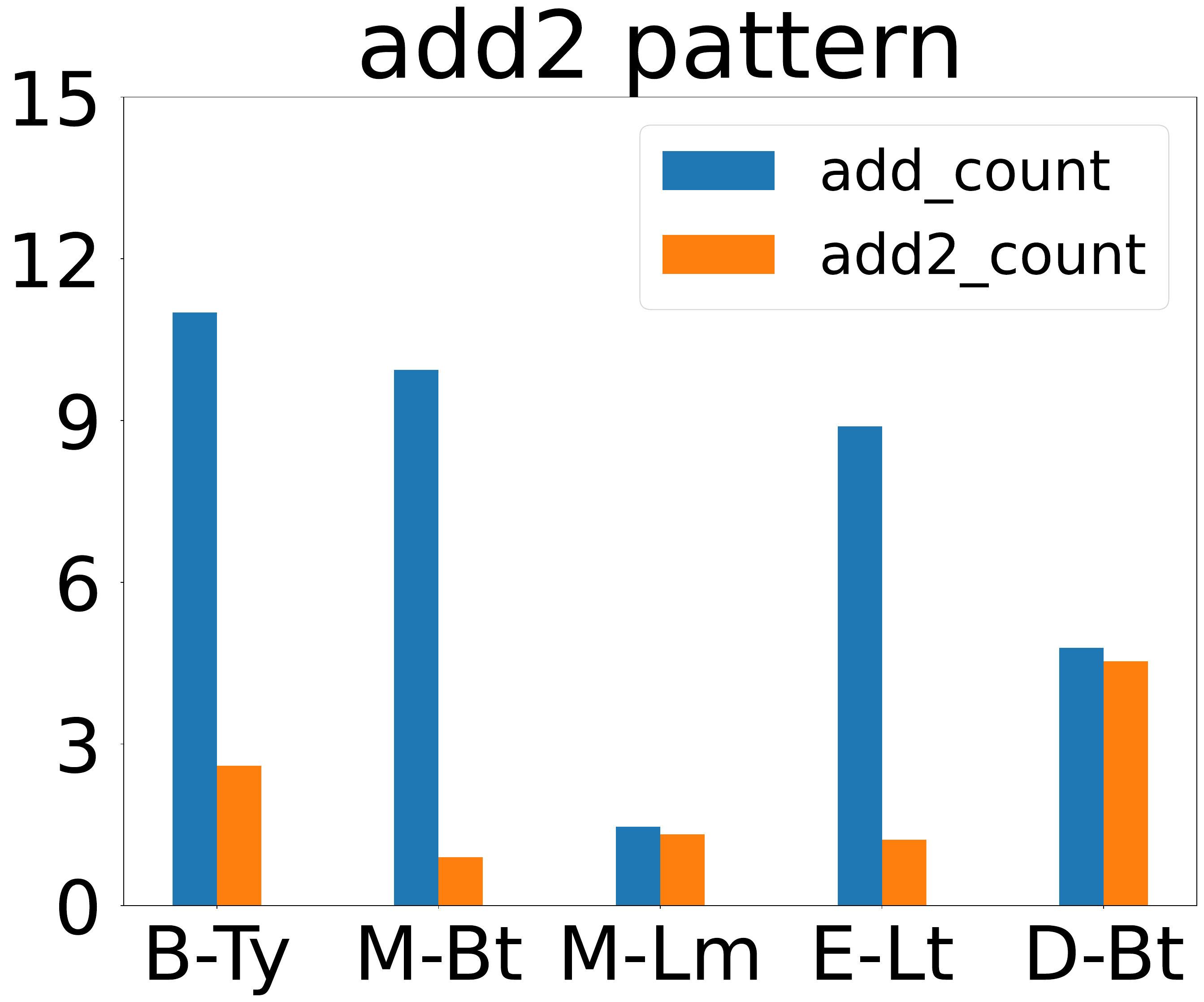}
  \end{minipage}
  \begin{minipage}{0.19\textwidth}
    \centering
    \includegraphics[height=2.8cm]{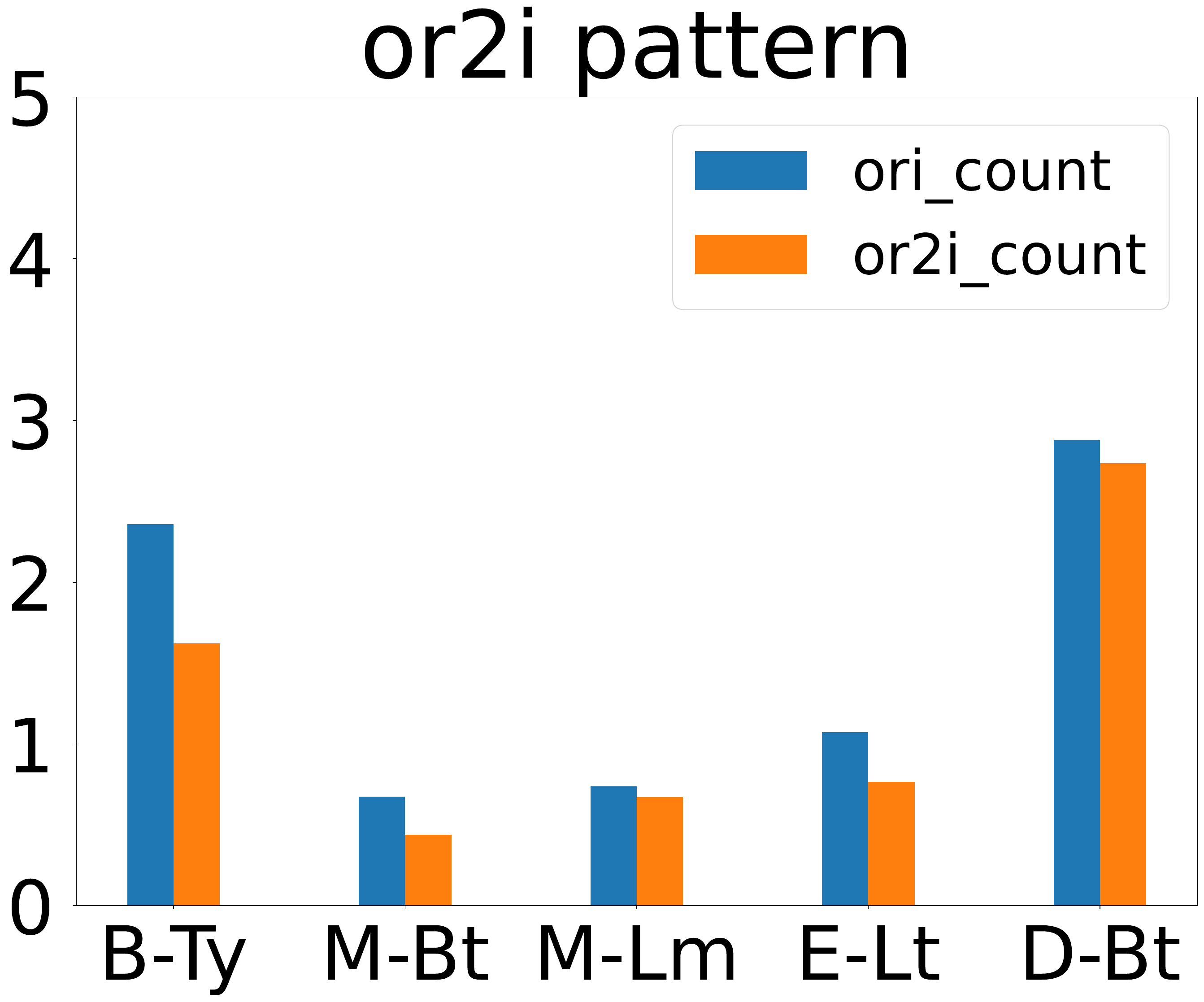}
  \end{minipage}
  \begin{minipage}{0.19\textwidth}
    \centering
    \includegraphics[height=2.8cm]{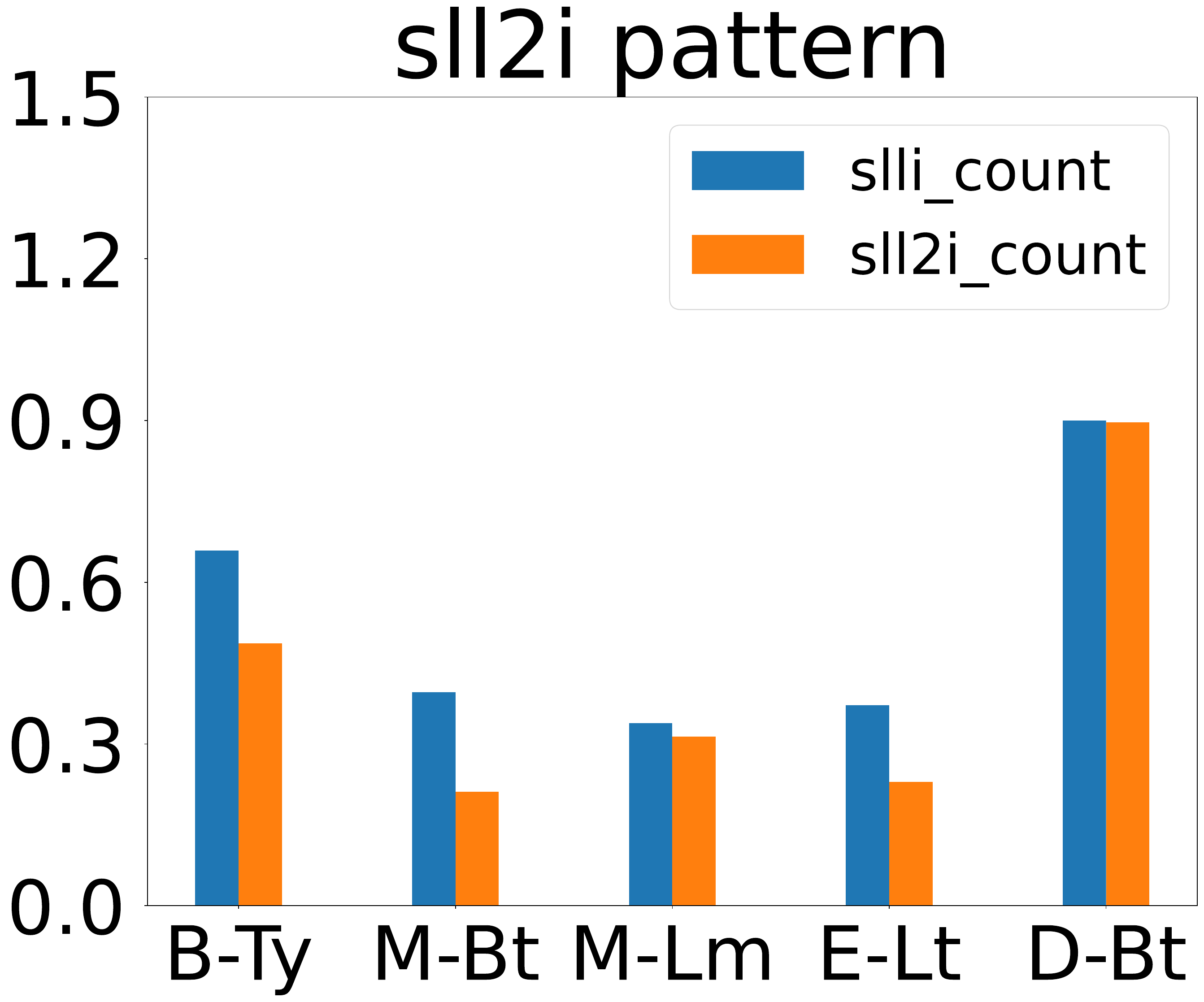}
  \end{minipage}%
  \caption{Frequently executed patterns identified through profiling of generated C code on baseline RISC-V}
  \label{pattern_count_plots}
  \vspace{-4mm}
\end{figure*}

\subsection{Model Preparation Phase}
Our flow targets encoder-only Transformer models. A pretrained Hugging Face model with a reinitialized classification head is fine-tuned on SST-2 dataset~\cite{socher-etal-2013-recursive}. The model is then exported to ONNX format and is subsequently compiled with TVM~\cite{10.5555/3291168.3291211}. Scripts are open-sourced\footnote{\url{https://github.com/ajaymahendersingh/Gen-RV-Tx}}.

\subsection{Compilation Phase}
\subsubsection{Profiling}
The TVM-generated C is compiled in ASIP Designer (version W-2024.12). The workloads are unquantized FP32. With RV32IM (no F-extension), the compiler realizes FP operations using software emulation libraries, severely degrading performance. Compiling for RV32IM\underline{\textbf{F}} maps FP32 operations to FPU instructions and removes soft-float libraries. The resulting binaries are executed on the Instruction Accurate (IA) simulator for profiling. Custom scripts extract hot instructions and patterns; the top ten by cycle count are examined. Beyond memory operations, several ALU operations (like \texttt{slli}, \texttt{add}, \texttt{addi}, \texttt{ori}) account for a significant share of cycles. Subsequently, an analysis of instruction sequences with high cycle costs was conducted. Fig.~\ref{pattern_count_plots} illustrates how much of the overall code is made up of these instructions and the extracted sequences, across models.

\subsubsection{ISA extensions}
We employ a data-driven methodology to propose targeted, and non-standard ISA extensions aimed at mitigating computational bottlenecks. Guided by profiling results, the baseline core is augmented with these custom extensions. Implemented using reserved/custom opcode space,
these extensions reuse the core’s existing general-purpose registers, no new architectural registers are added. The design and impact of these extensions are described below.

\par \textbf{add3i}:
Profiling shows frequent sequences of three independent \texttt{addi} instructions. We introduce \texttt{add3i}, which performs three parallel immediate additions in a single cycle using a shared 10-bit signed immediate broadcast to all operations. This width covers most observed cases in our evaluated models, capturing 94.76\% of B-Ty, 97.50\% of M-Bt, 96.72\% of M-Lm, 96.75\% of E-Lt, and 93.94\% of D-Bt cases. The instruction requires six registers; therefore, we reserve \texttt{x22}, \texttt{x23}, and \texttt{x30}.

\par \textbf{add2i}:
Profiling shows frequent back-to-back \texttt{addi} instructions. 
Across the evaluated models, one immediate operand in the instructions spans a substantially wider range than the other. We therefore encode one immediate in 10 bits and one in 5 bits — covering 96.92\% of B-Ty, 96.49\% of M-Bt, 86.97\% of M-Lm, 97.43\% of E-Lt, and 88.57\% of D-Bt cases. The instruction needs four registers, so we reserve registers \texttt{x22} and \texttt{x23}.

\par \textbf{add2}:
This instruction merges execution of two consecutive add instructions. Since there can be six distinct registers in two add instructions, we repurpose the RISC-V U-type instruction format for encoding. Four registers are mapped to the 20-bit immediate field, while the fifth register is fixed to (\texttt{x21)}. The sixth register is encoded in the source register field.

\par \textbf{or2i}:
This custom instruction is introduced to fuse two consecutive OR operations that each perform a bitwise OR with an immediate operand. Profiling reveals that the immediate values in these consecutive instructions remain identical across all evaluated models, with their magnitudes constrained below 1023. Therefore, a 10-bit field is sufficient to encode the immediate. Since the instruction requires encoding four registers but the available bit width is limited, one destination register is fixed to \texttt{x29}.

\par \textbf{sll2i}:
This custom instruction fuses two consecutive "shift left logical immediate (\texttt{slli})" operations. Since this instruction is encoded using the RISC-V I-type format, only two registers can be directly specified. To address this, we utilize the 12-bit immediate field to encode a third register, while the fourth register is fixed to \texttt{x24} in our design. The compiler automatically substitutes eligible sequences of consecutive \texttt{slli} instructions with \texttt{sll2i}, internally using \texttt{x24}.  On a 32-bit processor, the maximum shift amount is 31, requiring 5 bits to represent each immediate. Thus, two 5-bit fields are allocated within the instruction encoding.

\lstset{
    language={[x86masm]Assembler},
    basicstyle={\ttfamily\small},
    keywordstyle=\bfseries,
    morekeywords={sll2i, add2, add2i, zol},
    commentstyle=\itshape,
    numbers=left,
    numberstyle=\tiny,
    stepnumber=1,
    numbersep=5pt,
    tabsize=4,
    frame=single,
    columns=fullflexible,
    showstringspaces=false
}

\par
\begin{figure}
    \centering
    \includegraphics[height=4cm, width=1\linewidth]{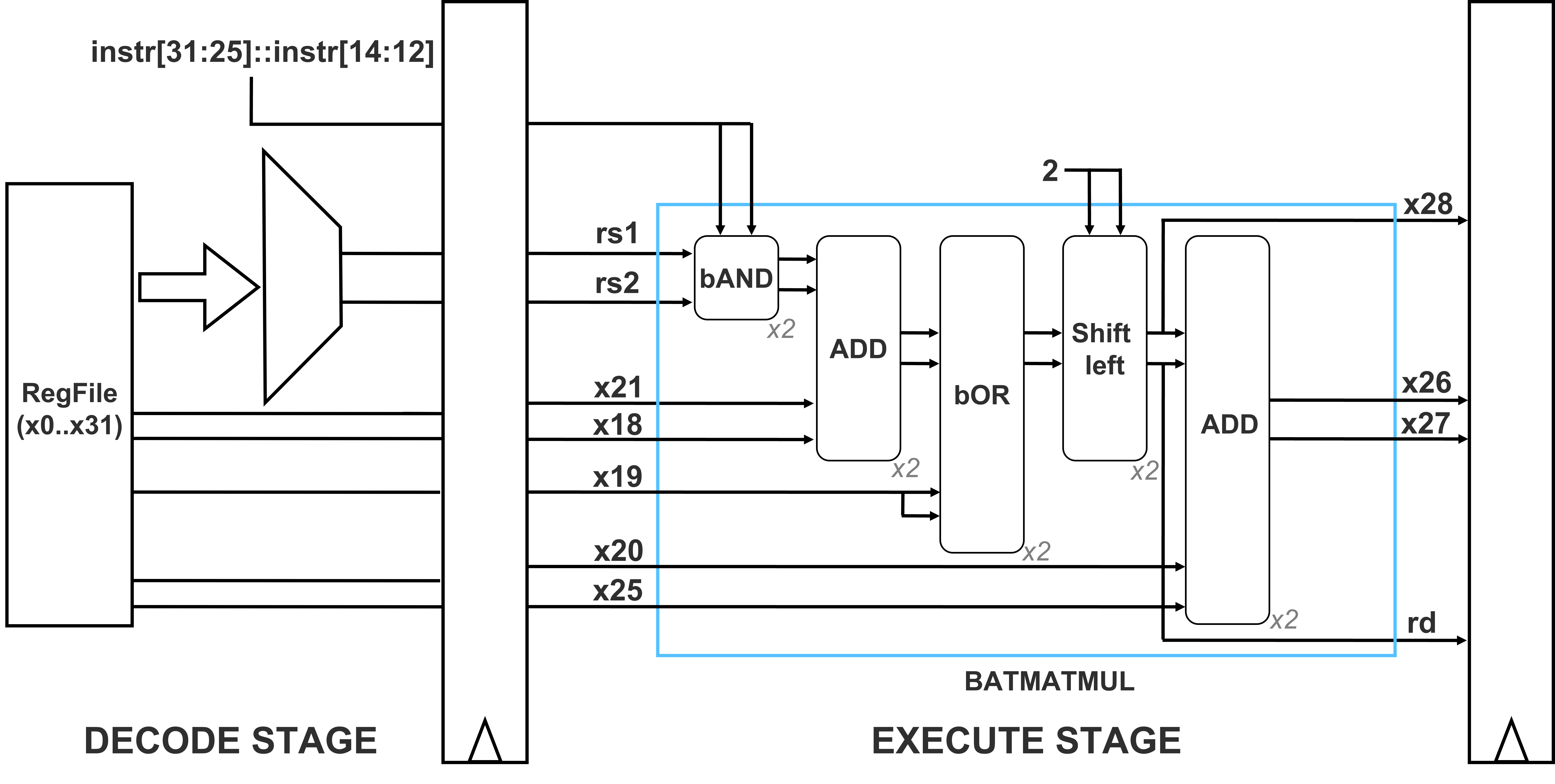}
    \caption{\texttt{batmatmul} HW}
    \label{batmatmul_HW}
    \vspace{-4mm}
\end{figure}

 \textbf{batmatmul}: Profiling identified batch matrix multiplications as one of the dominant consumers of clock cycles, motivating an atypical ISA extension to accelerate this workload. Although prior works like~\cite{10.1145/3658617.3697652} present high-performance matrix multiplication units, such designs incur significant area and power overhead. In contrast, we introduce a compact acceleration unit tightly integrated into the CPU pipeline that targets the address-generation stage of batch matrix multiplication. As illustrated in Fig.~\ref{batmatmul_HW}, the unit computes the parameter addresses that the floating-point unit subsequently fetches and multiplies, providing a practical trade-off between area, power and performance. Across the evaluated models, the proposed extension accelerates the batch matrix multiplication operation by 15.39–21.74\% on average. FPGA implementation shows hardware area and power overhead of 22.26\% and 4.42\%, respectively, while ASIC implementation shows 6.79\% area and 2.33\% power overheads. The microarchitecture compresses ten baseline instructions into a single fused operation. Guided by profiling of immediate values, two immediates are hardcoded and the remaining variable immediates are encoded in 10-bit fields; two of those variable immediates are identical, which further simplifies the encoding. Because the extension references eleven registers while encoding space is constrained, eight operand registers are fixed.
\vspace{-2mm}

\subsection{Hardware Generation Phase}
\begin{figure}
    \centering
    \includegraphics[height=4cm,width=1\linewidth]{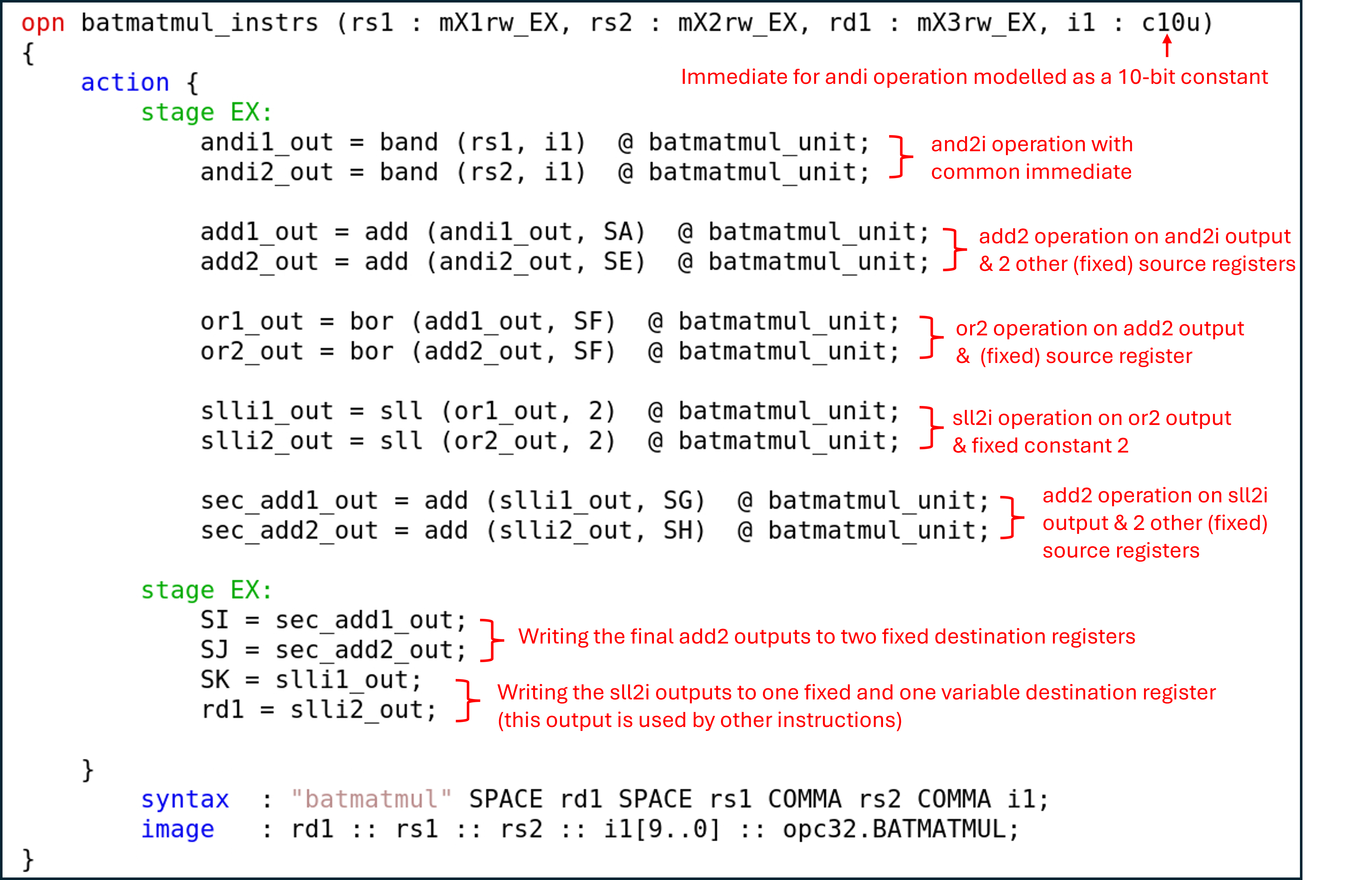}
    \caption{nML: \textit{batmatmul}}
    \label{nML_batmatmul}
    \vspace{-6mm}
\end{figure}

The obtained instruction set extensions are modeled in ASIP Designer using its domain-specific modeling language, nML. Fig.~\ref{nML_batmatmul} shows an example nML snippet implementing the proposed \textit{batmatmul} ISA extension. This is subsequently translated into synthesizable hardware. Given the significantly larger size of the models under consideration, functional validation of the generated RTL was performed via behavioral simulations. The functionality of the generated FPGA bitstream for AMD's ZCU102 FPGA platform has been validated through the execution of a small hand-coded transformer model, as described in \cite{10737828}.
\vspace{-2mm}

\subsection{Loop Optimization Strategy}
\label{loop_opt}
TVM-generated C code typically contains numerous \texttt{for} loops.
These loops translate to frequent branch instructions, which incur significant cycle overhead. Profiling shows that branch operations (\texttt{bne, beq}) are among the most cycle-intensive instructions, making loop optimization an important target. While MARVEL~\cite{11267280} demonstrates hardware loop extensions, this work explores a lightweight alternative using compiler-provided loop unrolling hints. Although this may increase code size, it requires no additional hardware, allows unrestricted unrolling, and provides fine-grained control to balance performance and memory usage. That said, hardware loop extensions remain preferable in scenarios where processor area constraints are less critical. Chess\textsuperscript{\textregistered} compiler allows users to annotate loops with unrolling hints. A custom script parses TVM-generated C codes and injects the required hints into all innermost \texttt{for} loops\footnote{except M-Bt, where the large number of \texttt{for} loops makes complete unrolling impractical and leads to compile-time issues}.

\section{Results}
This section presents the performance improvements and resource utilization associated with the extended trv32p3f core. 
\vspace{-4mm}
\subsection{Implementation}
FPGA resource utilisation for baseline and extended processor is shown in Table.~\ref{resource_util}. The extended core incurs a 32.94\% overall area overhead relative to the baseline. The overhead mainly arises from the decoder updates to accommodate new instructions. The usage of DSP blocks remains unchanged since the logic is implemented using CARRY8 FPGA primitives. Power consumption was estimated on the post-implementation design under typical process corner and default switching activity. The extended core shows an overhead of 7.23\%.

\begin{table}[h]
  \caption{FPGA resource utilisation of baseline and extended processor operating at 75 MHz}
  \label{resource_util}
  \resizebox{\columnwidth}{!}{%
    \begin{tabular}{lccccccc}
    \toprule
    \textbf{Processor}&CLB LUTs&CLB&Registers&CARRY8&MUX&DSP&Total Power (mW)\\
    \midrule
    \textbf{Baseline} & 9,026 & 1,816 & 2,168 & 128 & 1330 & 6 & 926\\
    \textbf{Extended} & 13,074 & 2,284 & 2,205 & 172 & 1501 & 6 & 993\\
    \midrule
  \textbf{Overhead:} & 4,048 & 468 & 37 & 44 & 171 & 0 & 67\\
  &  (44.84\%) & (25.77\%) & (1.70\%) & (34.37\%)& (12.85\%)&(0\%)&(7.23\%)\\
  \bottomrule
\end{tabular}
}
\vspace{-2mm}
\end{table}

ASIC synthesis (Table~\ref{asic_impl}) was also carried out for both the base and extended versions using Synopsys Design Compiler (version 2024.09) with TSMC 28 nm technology library at typical process corner, targeting a frequency of 250 MHz. 

\begin{table}[h]

\centering
  \caption{ASIC synthesis results using TSMC 28 nm PDK}
  \label{asic_impl}
    \begin{tabular}{lcc}
    \toprule
    \textbf{Processor}&Area (mm\textsuperscript{2}) & Power (mW)\\
    \midrule
    \textbf{Baseline} & 0.0166 & 0.4495\\
    \textbf{Extended} & 0.0201 & 0.4921\\
    \midrule
  \textbf{Overhead(\%):}&  21.08 & 9.47\\
  \bottomrule
\end{tabular}
\vspace{-4mm}
\end{table}

\subsection{Performance Evaluation}

Five LMs were benchmarked. All simulations were carried out using ASIP Designer, and the results obtained from the IA simulator are shown in Table.~\ref{cycle_and_inst_count}. The reported cycle count corresponds to the average number of cycles required to execute two inference runs. Note that \textcolor{blue}{trv32p3f\_ext\_U} refers to a processor variant that applies Chess compiler hints on top of the extended processor (see section~\ref{loop_opt}).

\begin{table}[h]
\vspace{-2mm}
    \centering
    \caption{Cycle Count (CC) and Instruction Count (IC) acceleration relative to baseline across LMs
    .}
    \label{cycle_and_inst_count}
    \resizebox{\columnwidth}{!}{%
    \begin{tabular}{lcccccccccc}
        \toprule
        \multirow{2}{*}{\textbf{Processor}} & \multicolumn{2}{c}{\textbf{B-Ty}} & \multicolumn{2}{c}{\textbf{M-Bt\textsuperscript{*}}} & \multicolumn{2}{c}{\textbf{M-Lm}} & \multicolumn{2}{c}{\textbf{E-Lt}} & \multicolumn{2}{c}{\textbf{D-Bt}} \\
        \cmidrule(lr){2-3} \cmidrule(lr){4-5} \cmidrule(lr){6-7} \cmidrule(lr){8-9} \cmidrule(lr){10-11}
        & \textbf{CC} & \textbf{IC} & \textbf{CC} & \textbf{IC} & \textbf{CC} & \textbf{IC} & \textbf{CC} & \textbf{IC} & \textbf{CC} & \textbf{IC} \\
        \midrule
        trv32p3f\_ext & 1.19 & 1.23 & 1.19 & 1.23 & 1.21 & 1.27 & 1.19 & 1.23 & 1.24 & 1.29\\
        \textcolor{blue}{trv32p3f\_ext\_U} & 2.13 & 1.88 & 1.50 & 1.47 & 2.10 & 1.80 & 2.19 & 1.91 & 2.07 & 1.81\\
        \bottomrule
    \end{tabular}%
    }
    \vspace{-3mm}
\end{table}

\begin{figure}[h]
  \centering
  \includegraphics[height=6cm,width=\linewidth]{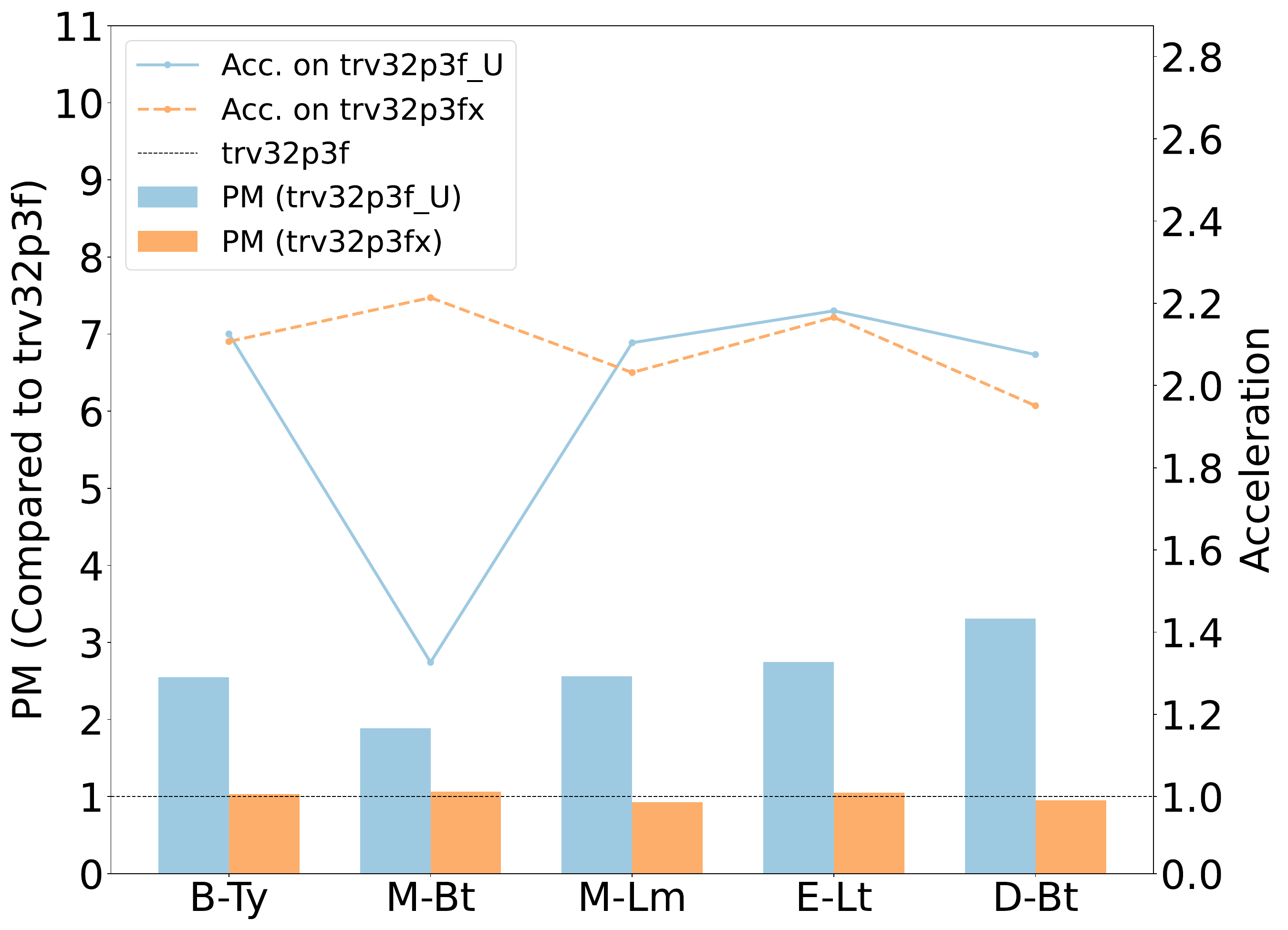}
    \put(-135,160){\scriptsize Note: M-Bt is not fully unrolled}
  \put(-120, 153){\scriptsize (see section~\ref{loop_opt})}
  \caption{Program Memory (PM) usage and cycle count for a single inference on cores with compiler loop-unrolling hints enabled (trv32p3f\_U) and hardware loop extensions (trv32p3fx), normalized with respect to the baseline (trv32p3f).}
  \label{trv32p3f_unrolled_vs_trv32p3fx}
  \vspace{-2mm}
\end{figure}
\vspace{-2mm}
\subsection{Software Loop Unrolling vs. Hardware Loop Extensions}

This section compares loop optimizations applied through Chess compiler hints 
against hardware-based loop acceleration. For the latter, we employ the Synopsys trv32p3fx core, which incorporates a hardware loop ISA extension, an address generation unit, and several DSP extensions. Although this core includes additional features beyond hardware loops, it provides a useful baseline for comparison. Fig.~\ref{trv32p3f_unrolled_vs_trv32p3fx} illustrates the variation in program memory usage and cycle counts between trv32p3f\_U and trv32p3fx, normalized with respect to trv32p3f, across the evaluated LMs. Incorporating compiler hints can more than double inference speed, making them ideal when program memory size is unrestricted and processor core area is tightly constrained.

\section{Conclusion and Future Work}
This paper introduces a framework for accelerating transformers, particularly encoder-only, on RISC-V. With the proposed extensions and compiler hints, inference speed more than doubles. Future work includes adding quantization to reduce C code memory footprint, and assessing a core without an FPU for higher speed operations.

\begin{small}
\bibliographystyle{IEEEtran}
\bibliography{thebibliography}
\end{small}

\end{document}